\documentclass[preprint,12pt]{elsarticle}

\usepackage{amssymb}
\usepackage{color}

\def \m {\mathrm}

\journal{Physica B: Condensed Matter}

\begin{document}

\begin{frontmatter}



\title{High-pressure Raman study on the superconducting pyrochlore oxide Cd$_2$Re$_2$O$_7$}


\author[a]{Yasuhito Matsubayashi}\ead{ymatsubayashi@issp.u-tokyo.ac.jp}
 \author[b]{Takumi Hasegawa} \author[b]{Norio Ogita} \author[c]{Jun-ichi Yamaura} \author[a]{Zenji Hiroi}\ead{hiroi@issp.u-tokyo.ac.jp}

\address[a]{Institute for Solid State Physics, The University of Tokyo, 5-1-5 Kashiwanoha, Kashiwa, Chiba 277-8581, Japan}
\address[b]{Graduate School of Integrated Arts and Sciences, Hiroshima University, 1-7-1 Kagamiyama, Higashi-Hiroshima, Hiroshima 739-8521, Japan}
\address[c]{Materials Research Center for Element Strategy, Tokyo Institute of Technology, Yokohama, Kanagawa 226-8503, Japan}

\begin{abstract}
The superconducting pyrochlore oxide Cd$_2$Re$_2$O$_7$ ($T_{\m{c}}=1\,{\m{K}}$), which is now considered as a candidate of the spin-orbit-coupled metal,
 shows an inversion-symmetry-breaking structural transition at $T_{\m{s1}} = 200 \,\m{K}$.
$T_{\m{s1}}$ decreases with increasing pressure and disappears at around $P_{\m{c}}=4.2 \,\m{GPa}$, where
 at least four high-pressure phases with tiny structural distortions are suggested by means of powder X-ray diffraction [Yamaura PRB 2017]. 
 We have carried out Raman scattering experiments to investigate changes in the crystal symmetry  under high pressures up to 4.8 GPa.
 A structural transition at 1.9-3.0 GPa and the recovery of inversion symmetry above $P_{\m{c}}$ are observed at 12 K.
\end{abstract}


\end{frontmatter}


\section{Introduction}
\label{introduction}

Cd$_2$Re$_2$O$_7$ is the only superconductor ($T_{\m{c}} = 1\,\m{K}$) in the family of pyrochlore 
oxides~\cite{hanawa2001superconductivity,sakai2001superconductivity,jin2001superconductivity}.
It shows two structural phase transitions,
from phase I ($Fd\bar{3}m$) to phase $\m{I\hspace{-.1em}I}$ ($I\bar{4}m2$) at 200 K ($T_{\m s1}$) with inversion symmetry breaking (ISB) and then 
to phase $\m{I\hspace{-.1em}I\hspace{-.1em}I}$ ($I4_122$) at 120 K ($T_{\m s2}$)~\cite{yamaura2002low}.
The phase transition at $T_{\m s1}$ is suggested to be driven by the band Jahn-Teller effect
because the density of states decreases largely below $T_{\m s1}$~\cite{hiroi2003correlations, vyaselev2002superconductivity}.
Probably related to this transition, a new idea of
"Spin-Orbit-Coupled Metal (SOCM)" has been proposed,
 which is a metal with inversion symmetry and a strong spin-orbit interaction~\cite{fu2015parity}.
In the SOCM, novel electronic orders such as a multipolar order can occur owing to the specific Fermi liquid instability
and are naturally expected to be accompanied by structural distortions with ISB.
Cd$_2$Re$_2$O$_7$ is one of candidates of the SOCM:
the phase transition at $T_{\m{s1}}$ is possibly induced by an electronic instability toward a multipolar phase.
Another interesting implication from the theory on the SOCM is the occurrence of 
odd-parity superconductivity mediated by parity fluctuations
in the vicinity of the quantum critical point of the ISB transition~\cite{kozii2015odd,wang2016topological}.
Probably corresponding to this, 
a large enhancement of the upper critical field was observed near $P_{\m{c}}=4.2 \,\m{GPa}$ in the previous high-pressure resistivity study~\cite{c2011superconductivity},
which suggests that an exotic superconductivity really takes place.

A very recent structural study under high pressures by means of powder x-ray diffraction suggests that the ISB transition is suppressed at around $P_{\m{c}}$. 
Moreover, 
at least four monoclinic phases [$\m{I\hspace{-.1em}V}$ ($Cc$), V ($C2/c$), $\m{V\hspace{-1pt}I\hspace{-1pt}I\hspace{-1pt}I}$ ($C2/m$), $\m{I\hspace{-1pt}X}$ ($Cm$)]
are observed near $P_{\m c}$
in addition to the three phases under ambient pressure~\cite{yamaura2017successive}.
However, partly because the observed lattice distortions are quite small, 
complete structural refinements could not be done, and the space groups were determined based on two assumptions.
One is to assume monoclinic systems for all the high-pressure phases, though the actual symmetries can be either monoclinic or triclinic.
The other assumption is the group-subgroup relationship for phase transitions: 
a second-order transition occurs within the same branch of symmetry lowering and a first-order transition between different branches. 
With these assumptions, the monoclinic space groups were uniquely decided to explain the experimental observations of second- and 
first-order transitions between them.
Thus, we think that more experimental information using other probes are needed to clarify the symmetry changes at high pressures.

In this study, we have carried out Raman scattering experiments under high pressures up to 4.8 GPa; Raman scattering
is a probe sensitive to a change in site symmetry, especially, to the presence of inversion symmetry.
The $T_{\m s1}$ transition of Cd$_2$Re$_2$O$_7$ at ambient pressure is related to the doubly degenerate $E_u$ irreducible representation
at the $\Gamma$ point in phase I~\cite{ishibashi2010structural,a2003structural}.
In phases $\m{I\hspace{-.1em}I}$ and $\m{I\hspace{-.1em}I\hspace{-.1em}I}$, this $E_u$ soft mode transforms into $A_1$ and $B_1$ modes.
Focusing on these modes, we
investigate pressure effects on Raman spectra at low temperature and clarify structural phase transitions and the recovery of inversion symmetry.

\section{Experimental}
\label{experiment}

Crystals of Cd$_2$Re$_2$O$_7$ were grown by the chemical vapor transport reaction.
High-quality crystals of Cd$_2$Re$_2$O$_7$ were grown through the improvement of synthesis conditions in the chemical vapor transport reaction, 
the details of which will be reported elsewhere~\cite{zenji}. The quality of the crystals was examined by the residual resistivity ratio (RRR) 
which is the ratio between resistivities at 300 and 2 K; the larger the RRR, the higher the crystal quality. 
The present crystals have larger RRR $\sim$ 60 than 20-30 for the previous crystals.
A polished crystal and a pressure media (methanol) were set into a gas membrane diamond anvil cell with a CuBe gasket.
The cell was pressurized by introducing He gas into the membrane after cooling to 12 K.
The pressure was measured by detecting the fluorescense spectra from ruby crystals inside the cell.
To estimate a possible pressure gradient in the cell,
the fluorescences from two ruby crystals were measured before each measurement of Raman spectra,
which showed a distribution of pressure of 0.2-0.4 GPa;
we employed an averaged value.
Measurements at 1 atm were conducted without the cell.
Raman spectra were excited with the 488 nm line of Ar-Kr laser [Stabilite 2017 (Spectra Physics)] with 8 mW power,
and a backscattered light was collected by a triple monochromator [NR-1800 (JASO)] and
a liquid-N$_2$ cooled CCD [LN/CCD-1340PB (Princeton Instruments Inc.)].

Raman spectra were measured on the polished (001) surface of a crystal.
Polarization measurements were conducted in two geometries: 
 (${\bf v}$,${\bf v}$) and (${\bf v}$,${\bf h}$) (${\bf v}//[100]$, ${\bf h}//[010]$),
where the first and second vectors in the bracket refer to the directions of the polarization direction of the incident
and the scattered lights, respectively.
In this report, 
the sum of the two spectra in (${\bf v}$,${\bf v}$) and (${\bf v}$,${\bf h}$) will be discussed
for high-pressure data, because unexpected rotations in the polarization direction occurred in the uniaxially compressed diamond.

\section{Results and Discussion}
\label{results}

\begin{figure}[tbp]
\begin{center}
\includegraphics[width=8cm,clip]{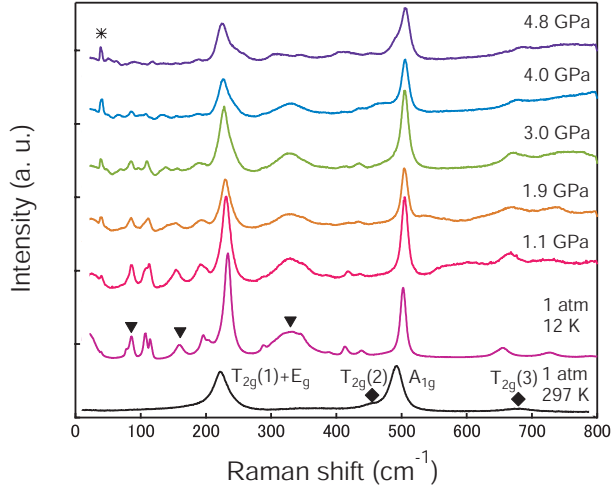}
\end{center}
\caption{Evolution of Raman spectra at 12 K with increasing pressure. 
Each spectrum is the sum of the two spectra measured in the (${\bf v}$,${\bf v}$) and (${\bf v}$,${\bf h}$) 
geometries with ${\bf v}//[100]$ and $h//[010]$.
A Raman spectrum recorded at 297 K and 1 atm is also shown at the bottom, in which most Raman-active peaks expected for
the $Fd\bar{3}m$ structure are observed:
$A_{1g}$ (492 $\m{cm^{-1}}$), $E_g$ (229 $\m{cm^{-1}}$), $T_{2g}$(1) (221 $\m{cm^{-1}}$), $T_{2g}(2)$ (451 $\m{cm^{-1}}$) and $T_{2g}$(3) (678 $\m{cm^{-1}}$);
the $E_g$ and $T_{2g}(1)$ peaks overlap each other and are identified separately by polarization measurements.
At 12 K and 1 atm, several additional peaks appear as a result of the ISB transition to phase $\m{I\hspace{-1pt}I\hspace{-1pt}I}$ ($I4_1 22$), which include
three $A_{1}^E$ modes marked by the inverted triangles.
The three pressure ranges of P1, P2 and P3 are noted at 1 atm-1.9 GPa, 3.0-4.0 GPa and 4.8 GPa, respectively.
The peak marked by $\ast$ is the natural emission from Ar$^{+}$ laser.
}
\label{all}
\end{figure}

First, the phase transitions of Cd$_2$Re$_2$O$_7$ at ambient pressure are considered.
Among the irreducible representations of phase I ($Fd\bar{3}m$) at the Brillouin zone center, which are
$
A_{1g}+3 A_{2u} + E_{g} + 3E_{u} + 2 T_{1g} + 8 T_{1u} + 4 T_{2g} + 4 T_{2u}
$,
six optical phonon modes ($A_{1g} + E_{g} + 4 T_{2g}$) are Raman-active.
In  the spectrum of phase I measured at 297 K and 1 atm
 (Fig. \ref{all}), five Raman peaks are assigned by polarization measurements: 
$A_{1g}$ (492 $\m{cm^{-1}}$), $E_g$ (229 $\m{cm^{-1}}$), $T_{2g}$(1) (221 $\m{cm^{-1}}$), $T_{2g}$(2) (451 $\m{cm^{-1}}$), $T_{2g}$(3) (678 $\m{cm^{-1}}$).
One missing $T_{2g}$ peak may be too weak to observe, which was also the case in the previous report~\cite{kendziora2005goldstone}.
The $E_g$ and $T_{2g}(1)$ modes overlap each other but are identified separately in different polarization measurements.

The irreducible representations of phase $\m{I\hspace{-1pt}I\hspace{-1pt}I}$ ($I4_1 22$) are enormously increased as
$
5A_1 + 10A_2 +8B_1 +7B_2 + 18E
$.
The number of optical Raman modes also increases to 37 ($5A_1 + 8B_1 + 7B_2 + 17E$),
compared to phase I.
Note that the acoustic modes are $A_2 + E$.
19 Raman peaks are observed in
the spectrum measured at 12 K and 1 atm for phase $\m{I\hspace{-1pt}I\hspace{-1pt}I}$ in
Fig. \ref{all}.
The rest of the Raman-active peaks may
 be weak or overlap with others.
The detail of peak assignments will be reported elsewhere~\cite{matsubayashi}.

The soft mode associated with the phase transition at $T_{\m{s1}}$ at ambient pressure is
Raman-inactive $E_u$ mode in phase I,
which transforms into the Raman-active $A_1$ and $B_1$ modes in phase $\m{I\hspace{-1pt}I\hspace{-1pt}I}$.
This is the consequence of ISB, and, thus, the presence of the two modes is a clear signature of ISB.
In phase I, 
3 $E_u$ modes which include the soft mode of the $T_{\m{s1}}$ transition exist, while
in phase $\m{I\hspace{-1pt}I\hspace{-1pt}I}$
5 $A_1$ and 8 $B_1$ modes are allowed, which include 3 $A_1$ and 3 $B_1$ modes derived from the 3 $E_u$ modes in phase I.
In this paper the 3 $E_u$-derived $A_1$ modes are focused, which are labeled as the $A_1 ^E$ modes.
They are observed in the spectrum at 12 K and 1 atm shown in Fig. \ref{all}: $A_1 ^E (1)$ (86 $\m{cm^{-1}}$), $A_1 ^E (2)$ (159 $\m{cm^{-1}}$),
 and $A_1 ^E (3)$ (335 $\m{cm^{-1}}$).
Among the three $A_1 ^E$ modes, the soft mode of the $T_{\m{s1}}$ transition
has been assigned to $A_1 ^E$(1) at 85 cm$^{-1}$ at 6 K
in the previous report~\cite{kendziora2005goldstone}.

\begin{figure}[tbp]
\begin{center}
\includegraphics[width=13cm,clip]{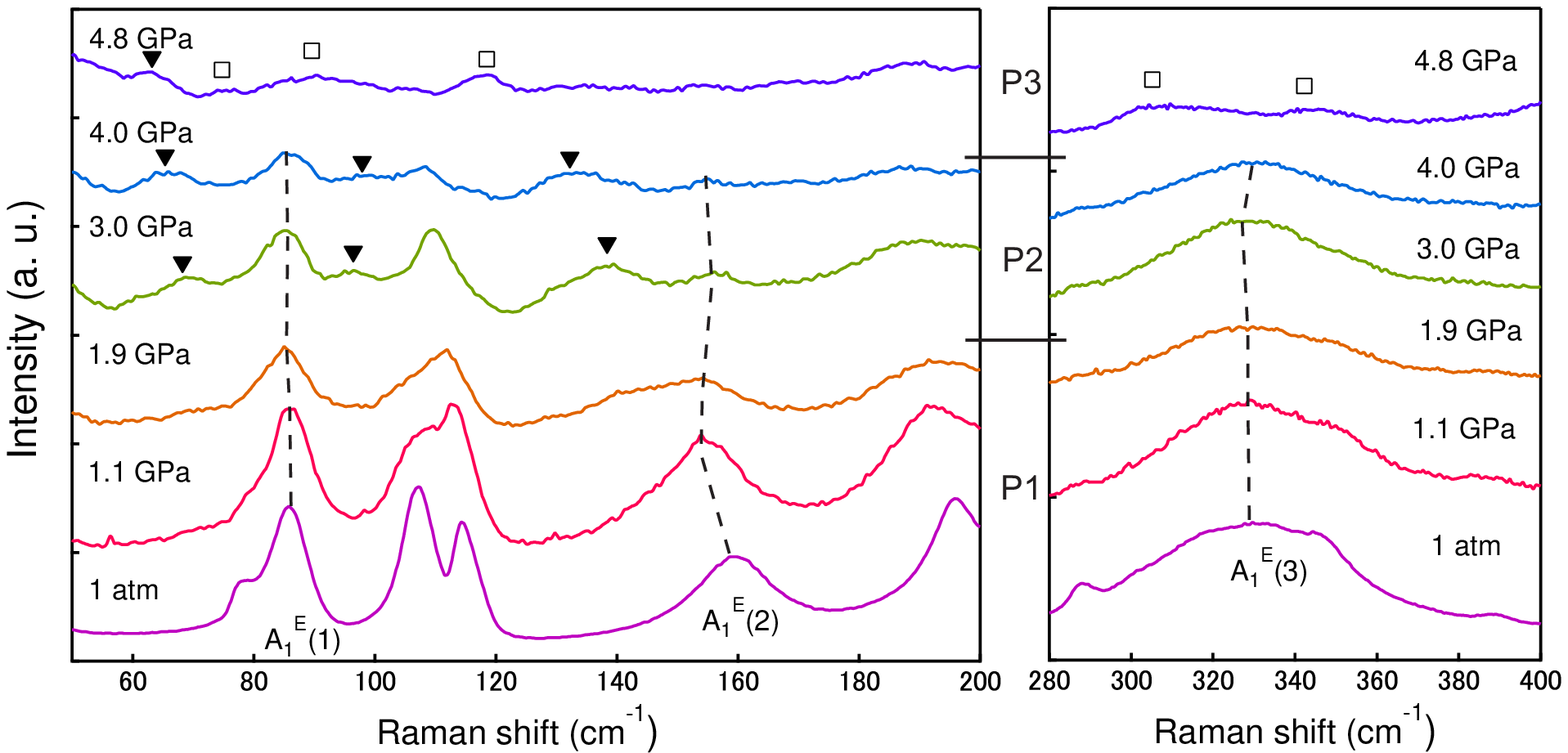}
\end{center}
\caption{Enlargements of the Raman spectra of Fig. \ref{all} at 50-200 $\m{cm^{-1}}$ (left) and 280-400 $\m{cm^{-1}}$ (right).
The evolutions of the three $A_1 ^E$ modes originating from the $E_u$ mode of the $Fd\bar{3}m$ structure are shown by the broken lines: 
$A_{1}^{E}\m{(1)}$, $A_{1}^{E}\m{(2)}$, $A_{1}^{E}\m{(3)}$ at 86, 159 and 336 cm$^{-1}$ at 12 K and 1 atm, respectively.
The three peaks marked by inverted triangles at 3.0 and 4.0 GPa are new peaks appearing possibly due to the phase transition to phase $\m{I\hspace{-.1em}V}$.
The disappearance of the $A_1 ^E$ modes and the
appearance of the five peaks marked by squares in the 4.8 GPa spectrum may correspond to the ISB transition to phase $\m{V\hspace{-1pt}I\hspace{-1pt}I\hspace{-1pt}I}$.
}
\label{soft}
\end{figure}

Now we describe the evolution of Raman spectra at 12 K with increasing pressure (Fig. \ref{all}). 
Three types of spectra are obtained
in the pressure ranges of
 P1 (1 atm, 1.1 and 1.9 GPa), P2 (3.0 and 4.0 GPa) and P3 (4.8 GPa).
Subtle but significant changes in the spectra are noticed between the three groups by
focusing the $A_1 ^E$ modes in Fig. \ref{soft}.

As pressure increases, the spectra remain essentially the same up to 1.9 GPa in group P1,
while
three new peaks  are observed at 69, 96 and 138 cm$^{-1}$ at 3.0 GPa,
and they move to 66, 98 and 132 cm$^{-1}$ at 4.0 GPa, respectively.
This evidences that a structural phase transition occurs between 1.9 and 3.0 GPa.

Another distinct change is observed between 4.0 and 4.8 GPa in Fig. \ref{soft},
where the intensities of the 3 $A_1 ^E$ modes decrease dramatically and almost vanish.
This indicates the recovery of inversion symmetry.
In addition, five new peaks, which are marked by the squares in Fig. \ref{soft}, 
appear at 76, 89, 118, 305 and 342 cm$^{-1}$;
the peak at 63 cm$^{-1}$, marked by the inverted triangle, is considered to
correspond to those at
69 cm$^{-1}$ at 3.0 GPa and 66 cm$^{-1}$ at 4.0 GPa.
These results indicate that there is a structural transition accompanied by the recovery of inversion symmetry at 4.0-4.8 GPa.

Now we would like to discuss the phase transitions at 12 K between 1 atm and 4.8 GPa
in comparison with the previous results based on resistivity and XRD measurements.
Concerning the phase transition from phase $\m{I\hspace{-1pt}I\hspace{-1pt}I}$ ($I4_1 22$) to phase $\m{I\hspace{-1pt}X}$ ($Cm$),
 which suggested to occur at 0-1 GPa,
our Raman experiments do not detect any change in the spectra at the corresponding pressure region.
This suggests that the transition causes a really tiny distortion or does not exist.
When entering to group P2, the additional peaks appear without any peaks disappearing;
such behavior is often observed in a symmetry-lowering transition to a subgroup.
This change must correspond to the transition from phase $\m{I\hspace{-1pt}X}$ ($Cm$) or phase $\m{I\hspace{-1pt}I\hspace{-1pt}I}$ ($I4_1 22$) 
to phase $\m{I\hspace{-1pt}V}$ ($Cc$) found at 2.8 GPa.
However, the $Cm$-to-$Cc$ and $I4_1 22$-to-$Cc$ transitions which do not satisfy a group-subgroup relation
seem to be incompatible with our observation. Thus, there is a reason to reconsider the proposed space groups.
On the other hand, in group P3, the three $A_1 ^E$ modes disappear with the new peaks appearing,
which suggests a transition without a group-subgroup relation.
This is consistent with the 1st order transition from phase $\m{I\hspace{-1pt}V}$ ($Cc$) to phase $\m{V\hspace{-1pt}I\hspace{-1pt}I\hspace{-1pt}I}$ ($C2/m$).
More importantly, both the present and previous studies consistently indicate 
the recovery of inversion symmetry above $P_{\m{c}} = 4.2 \,\m{GPa}$.

It is known that the polar-nonpolar transition in ferroelectric materials is suppressed by applying pressure,
as observed in Cd$_2$Re$_2$O$_7$.
This is because short-range interactions that stabilize the centrosymmetric structure become relatively stronger than
long-range dipole-dipole interactions that stabilize the relevant soft phonon mode, as discussed by Samara {\it et al.}~\cite{samara1982study}.
This Samara's rule may not simply apply to Cd$_2$Re$_2$O$_7$, because the Coulomb interaction is effectively screened by conduction electrons 
in such a metallic compound. Therefore, we should consider an alternative mechanism for the suppression of the ISB transition. 
The Fermi liquid instability of the SOCM may be a possible origin.

In the SOCM, an ISB structural transition 
can be induced by a
strong coupling between the electronic instability and odd-parity phonons,
which result in an odd-parity multipolar order.
Based on the Landau theory, the order parameter of the ISB transition of Cd$_2$Re$_2$O$_7$ is considered to be 
the doubly degenerate $E_u$ irreducible representation at the $\Gamma$ point in phase I~\cite{ishibashi2010structural,a2003structural}.
On the other hand, the recent second harmonic optical anisotropy measurements claim that the order parameter 
of the transition is 
the $T_{2u}$ representation~\cite{harter2017parity};
the $E_u$ representation is suggested to be a secondary one.
Anyway, there may be various routes to release the Fermi liquid instability
which may lead to a number of multipolar orders
 in Cd$_2$Re$_2$O$_7$.
Provided that the ISB transition at ambient pressure is caused by the $E_u$ soft modes, the
two noncentrosymmetric phases of $\m{I\hspace{-1pt}X}$ ($Cm$) and $\m{I\hspace{-1pt}V}$ ($Cc$) under high pressures
may be derived from other odd-parity representations such as $A_{2u}$, $T_{1u}$ and $T_{2u}$.
Moreover, the high-pressure phases with inversion symmetry, which are phase V ($C2/c$), phase $\m{V\hspace{-1pt}I\hspace{-1pt}I\hspace{-1pt}I}$ ($C2/m$)
and the $R\bar{3}m$ phase above 21 GPa~\cite{malavi2016cd}, are possibly induced by electronic instabilities coupled with even-parity phonons towards even-parity multipolar orders.
Therefore, very rich physics on the interplay between the electronic instability and crystal structure of the SOCM must be involved in Cd$_2$Re$_2$O$_7$. 
Note that all the low-temperature phases are rendered superconducting at low temperatures possibly with different characters reflecting 
the corresponding multipolar orders. For further information, it is crucial to decide the symmetry of the crystal structures of the multipolar phases. 
Future Raman scattering experiments would clarify the mechanism of these phase transitions through the determination of relevant phonon modes.

\section{Conclusion}

Structural changes of Cd$_2$Re$_2$O$_7$ under high pressures up to 4.8 GPa 
have been examined by Raman scattering experiments at 12 K.
The $A_1$ modes derived from the $E_u$ modes of phase I disappear at 4.0-4.8 GPa,
which corfirms the recovery of inversion symmetry
above $P_c = 4.2 \,\m{GPa}$ reported in the previous XRD study.
Moreover, the transition to phase $\m{I\hspace{-1pt}V}$ is confirmed by the appearance of new Raman peaks at 1.9-3.0 GPa.
 On the other hand, we do not
observe any changes in the Raman spectra 
below 1.9 GPa and at 3.0-4.0 GPa, which suggests the absence of the transitions to 
phases $\m{I\hspace{-1pt}X}$ and $\m{V\hspace{-1pt}I\hspace{-1pt}I}$ reported in the previous study, respectively.
Further Raman scattering experiments under high pressures 
and in a wide temperature range are undergoing, which will 
clarify the whole phase diagram of
 Cd$_2$Re$_2$O$_7$.

\section*{Acknowledgements}

Y.M. is supported by the Materials Education
Program for the Future Leaders in Research, Industry, and
Technology (MERIT) given by the Ministry of Education,
Culture, Sports, Science and Technology of Japan (MEXT).
This work was partially supported by the Core-to-Core Program for Advanced
Research Networks given by the Japan Society for the
Promotion of Science (JSPS).

\bibliographystyle{elsarticle-num} 
\bibliography{CROraman}

\end{document}